 \documentclass[final,1p,times]{elsarticle}

\usepackage{amssymb}

\usepackage{url}

\journal{Nuclear Physics A}

\begin{document}

\begin{frontmatter}



\long\def\symbolfootnote[#1]#2{\begingroup%
\def\thefootnote{\fnsymbol{footnote}}\footnote[#1]{#2}\endgroup}

\title{X-ray transition yields of low-$Z$ kaonic atoms produced in Kapton}


\author[lnf]{M.~Bazzi}
\author[victoria]{G.~Beer}
\author[lnf,smi]{C.~Berucci}
\author[milano]{L.~Bombelli}
\author[lnf,ifin]{A.M.~Bragadireanu}
\author[smi]{M.~Cargnelli}
\author[lnf]{C.~Curceanu}
\author[lnf]{A.~d'Uffizi}
\author[milano]{C.~Fiorini}
\author[roma]{F.~Ghio}
\author[lnf]{C.~Guaraldo}
\author[ut]{R.S.~Hayano}
\author[lnf]{M.~Iliescu}
\author[smi]{T.~Ishiwatari\corref{cor1}}
\ead{tomoichi.ishiwatari@assoc.oeaw.ac.at}
\cortext[cor1]{T.~Ishiwatari}
\author[riken]{M.~Iwasaki}
\author[tum]{P.~Kienle\symbolfootnote[2]{deceased}}
\author[lnf]{P.~Levi Sandri}
\author[milano]{A.~Longoni}
\author[smi]{J.~Marton}
\author[riken]{S.~Okada}
\author[lnf,ifin]{D.~Pietreanu}
\author[ifin]{T.~Ponta}
\author[milano]{R.~Quaglia}
\author[Santiago]{A.~Romero~Vidal}
\author[lnf]{E.~Sbardella}
\author[lnf]{A.~Scordo}
\author[ut]{H.~Shi}
\author[lnf,ifin]{D.L.~Sirghi}
\author[lnf,ifin]{F.~Sirghi}
\author[lnf]{H.~Tatsuno\corref{cor2}}
\ead{hideyuki.tatsuno@lnf.infn.it}
\cortext[cor2]{H.~Tatsuno}
\author[ifin]{A.~Tudorache}
\author[ifin]{V.~Tudorache}
\author[lnf,tum]{O.~Vazquez~Doce}
\author[smi]{E.~Widmann}
\author[smi]{J.~Zmeskal}
\author[]{(SIDDHARTA~collaboration)}

\address[lnf]{ INFN, Laboratori Nazionali di Frascati, C.P. 13, Via E. Fermi 40, I-00044 Frascati (Roma), Italy}
\address[victoria]{Department of Physics and Astronomy, University of Victoria, P.O. Box 1700 STN CNC, Victoria BC V8W 2Y2, Canada}
\address[smi]{Stefan-Meyer-Institut f\"{u}r subatomare Physik, Boltzmanngasse 3, 1090 Wien, Austria}
\address[milano]{Politecnico di Milano, Dipartimento di Elettronica e Informazione, Piazza L. da Vinci 32, I-20133 Milano, Italy}
\address[ifin]{IFIN-HH, Institutul National pentru Fizica si Inginerie Nucleara Horia Hulubei, Reactorului 30, Magurele, Romania}
\address[roma]{INFN Sezione di Roma I and Instituto Superiore di Sanita, I-00161 Roma, Italy}
\address[ut]{University of Tokyo, 7-3-1, Hongo, Bunkyo-ku, Tokyo, Japan}
\address[riken]{RIKEN, Institute of Physical and Chemical Research, 2-1 Hirosawa, Wako, Saitama 351-0198, Japan}
\address[tum]{Excellence Cluster Universe, Technische Universit\"{a}t M\"{u}nchen, Boltzmannstra{\ss}e 2, D-85748 Garching, Germany}
\address[Santiago]{Universidade de Santiago de Compostela, Casas Reais 8, 15782 Santiago de Compostela, Spain}
\begin{abstract}
The X-ray transition yields of kaonic atoms produced in Kapton polyimide
($\rm{C}_{22}\rm{H}_{10}\rm{N}_{2}\rm{O}_{5}$) were measured for the first time in the  
SIDDHARTA experiment. X-ray yields of the kaonic atoms with low atomic numbers 
($Z = 6,7,$ and 8) and  transitions with high principal quantum numbers ($n = 5 - 8$) were determined. 
The relative yields of the successive transitions in the same atoms and 
the yield ratios of carbon-to-nitrogen (C:N) and carbon-to-oxygen (C:O) for the same transitions
were also determined. These X-ray yields provide important information for 
understanding the capture ratios and cascade mechanisms of 
kaonic atoms produced in a compound material, such as Kapton.
\end{abstract}

\begin{keyword}
 kaonic atoms \sep
X-ray yields \sep
 cascade processes \sep
 X-ray detection
\end{keyword}

\end{frontmatter}


\section{Introduction}
\label{sec:intro}

The study of kaonic atoms provides unique information on low-energy QCD in the strangeness sector.
The energy shifts and widths of the lowest-lying levels permit extraction of 
the strong interaction in the low-energy regime. See, for example, the recent experimental results 
\cite{sidd-kh,sidd-khe4,sidd-khe3,sidd-khewidth}, and references therein.
The upper-level widths caused by the strong interaction can be determined using the relative yields
of the lowest-lying levels and the levels feeding it.
These widths have been used for theoretical calculations
of the kaon-nucleon/nucleus interaction (e.g., \cite{upper,exa-friedman,npa-friedman,npa-friedman2}). 

On the other hand, even if higher X-ray transitions are not significantly affected by the 
strong interaction between the kaon and nucleus, 
the X-ray yields provide fundamental information on the atomic capture, 
initial distribution and de-excitation processes.

Cascade calculations based on kaonic atom X-ray yields taken with several solid targets 
\cite{Wiegand69,WG} were initially performed using the Fermi-Teller model \cite{leon-seki}.
The results showed a smooth variation of the yields  over atomic number $Z$, as shown in Fig.~2 
of Ref. \cite{leon-seki}.

When this data set was extended to targets ranging from helium ($Z=2$) through 
uranium ($Z=92$) \cite{Wiegand-Godfrey74}, a striking result was obtained. 
Unlike the earlier prediction of Ref.~\cite{leon-seki}, an unexpected variation 
in the pattern of X-ray yields vs atomic number was observed. This variation has 
several maxima and minima, as shown in Fig. 7 of Ref.~\cite{Wiegand-Godfrey74}, 
where the maxima occur near the closed atomic electron shells.

Similar yield patterns were observed in muonic and pionic atom X-ray yields, as well as
positron-annihilation lifetimes in annealed metals \cite{Kunselman}. 
Theoretical studies showed that these patterns could be related to the 
properties of the exotic atoms, such as the atomic radii, the initial distributions of the 
particles, etc. \cite{Kunselman,Condo}. 

Although a large amount of data on kaonic atom X-ray yields was collected, transitions
with a high principal number $n$ in materials of low atomic number $Z$
have not previously been measured due to the low energy of the X-rays in such transitions. 
It is of interest to explore such unmeasured regions both experimentally and theoretically. 

Previous yield measurements were mainly of pure solid materials. 
However, X-ray yields in compounds such as hydrides (LiH, CH, NaH, and CaH$_{2}$)
were reported to have reductions of 20-70\% \cite{PLB-WG,Wiegand-Lum}, 
possibly related to the Stark effect of the protons in the hydrides and the kaon transfers. 

In this paper, we report on the first determination of the X-ray yields of 
kaonic atoms produced in the Kapton polyimide ($\rm{C}_{22}\rm{H}_{10}\rm{N}_{2}\rm{O}_{5}$) foils, 
which were used as a target window material in the SIDDHARTA experiment. 

\section{The SIDDHARTA experiment}
\label{sec:exp}

The SIDDHARTA experiment was performed using low-energy kaons produced by the DA$\Phi$NE 
electron-positron collider. The main purpose of the experiment was the measurement of
X-rays from kaonic atoms formed in gaseous targets. In addition, we measured kaonic atom X-rays produced 
in the windows of the gaseous target cell. These latter X-ray yields are reported in this paper.

\begin{figure}
\begin{center}
\includegraphics[width=0.6\linewidth]{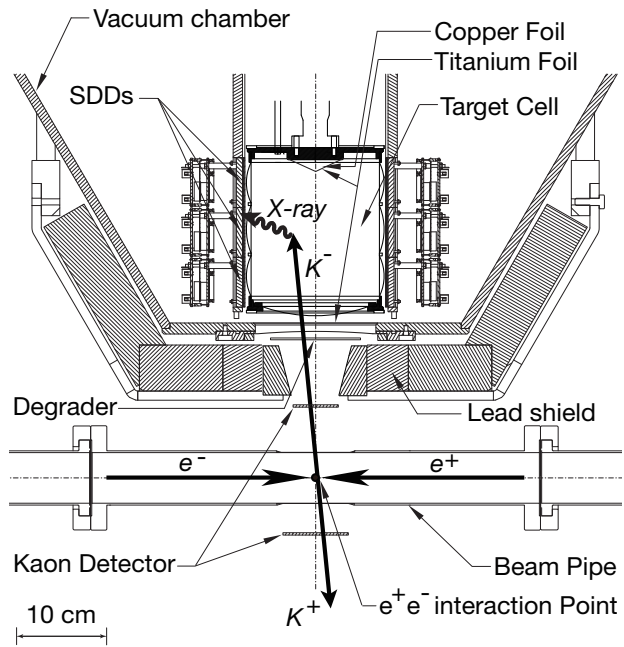}
  \caption{A schematic view of the SIDDHARTA setup installed at
  the e$^+$e$^-$ interaction region of DA$\rm{\Phi}$NE.}
  \label{fig:Setup}
\end{center}
\end{figure}

Figure \ref{fig:Setup} shows the schematic view of the experimental setup. 
Electrons and positrons collided at the DA$\Phi$NE interaction point, where almost at-rest
$\phi$ mesons were created. $K^+K^-$ pairs produced by $\phi$ decay were emitted 
from the interaction point. The $K^+K^-$ pairs were detected by a kaon detector consisting 
of two plastic scintillators placed above and below the interaction point. 
The timing signals of the $K^+K^-$ pairs were recorded using  clock signals with 
a frequency of 120 MHz. Because of a finite crossing angle of the electron and position beams,
the momentum of the emitted kaons had an angular dependence. This angular dependence was compensated 
using a stepped degrader made of Mylar foils with thicknesses ranging from 100 to 800 $\mu$m. 
The thicknesses of the degrader were optimized to increase kaon stops in the target gas.

The charged kaons entered the target cell after passing through the scintillator and degrader.
The cylindrical target cell, made of Kapton polyimide ($\rm{C}_{22}\rm{H}_{10}\rm{N}_{2}\rm{O}_{5}$) 
foils with a thickness of 75 microns and a density of 1.42 g/cm$^3$, was 
15.5 cm high with a diameter of 13.7 cm.
The bottom of the target cell, through which the charged kaons entered, was also made of Kapton.
On the top of the target cell, thin Ti and Cu foils were installed to 
produce fluorescence X-rays induced by the beam background. 

In the experiment, we used four target gases: hydrogen (1.30 g/$l$), deuterium (2.50 g/$l$), 
helium-3 (0.96 g/$l$) and helium-4 (1.65 g/$l$ and 2.15 g/$l$). They were cooled to 23 K. 
The charged kaons stopped mainly in the target gas and partially in the Kapton windows.
The number of kaons stopping in the target gas and  the windows, which depends on the target density, 
was determined by  Monte Carlo simulations.

\begin{figure}
\begin{center}
\includegraphics[width=0.8\linewidth]{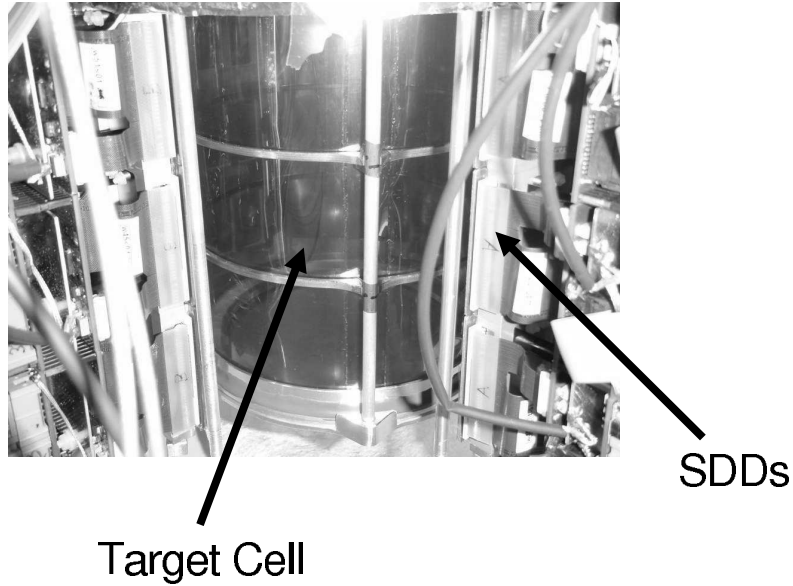}
  \caption{A picture of the target cell made of Kapton and the SDDs together with the readout electronics.}
  \label{fig:target}
\end{center}
\end{figure}

The X-rays were measured by silicon drift detectors (SDDs), 
which surrounded the target as shown in Fig.~\ref{fig:target}.
 Each SDD has an effective area of 1 cm$^2$ with a 450-$\mu$m thickness.
There were 144 SDDs in total. The SDDs were cooled to a temperature of 170 K to obtain an excellent 
energy resolution of about 150 eV for 6 keV X-rays, which is close to the resolution in noise-free 
conditions.

The data on the kaonic atom X-rays were accumulated in 2009. The X-ray yields of the kaonic atoms 
produced in the Kapton windows were determined using the data taken with deuterium gas
(an integral luminosity of 92 pb$^{-1}$). In these data, since kaonic deuterium 
X-ray peaks were very small~\cite{kd}, X-ray peaks from Kapton can be easily
extracted. In contrast, the data taken with other gases were not suitable due to overlapping peaks.

\section{Data analysis}
\label{sec:ana}

The data taken with the SDDs and the kaon detector were analyzed to extract kaonic atom X-ray spectra.
The analysis methods given in Refs.~\cite{sidd-khe3,sidd-khewidth} were used 
in the following procedures with the addition of a method dedicated to determination of the X-ray yields.

\begin{figure}
\begin{center}
\includegraphics[width=0.6\linewidth]{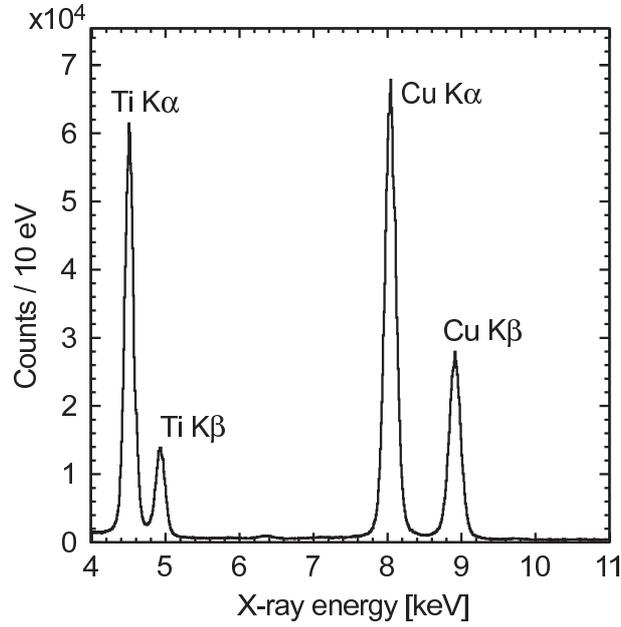}
 \caption{A typical energy spectrum for one SDD in the calibration data
with the X-ray tube irradiating the Ti and Cu foils.
The energy scale was calibrated using the positions of the Ti and Cu K$\alpha$ lines. 
}
 \label{calib}
\end{center}
\end{figure}

The energy scales were calibrated for each SDD, where the data taken with the Ti and Cu foils 
excited by an X-ray tube and the beam background were used \cite{sidd-kh,sidd-khe4,sidd-khe3,sidd-khewidth,kd}. 
A typical energy spectrum of the data taken with one SDD is shown in Fig.~\ref{calib}. High statistics 
of the fluorescence Ti and Cu X-rays were obtained. Since the X-ray energies of these lines
are well known \cite{ene1,ene2}, the energy scale was calibrated using 
the peak positions of the K$\alpha$ lines, assuming good integral linearity of 
the analogue-to-digital converters (ADCs). The Ti and Cu K$\beta$ lines are not suitable for a 
detailed energy calibration due to the existence of  satellite transitions 
close to the K$\beta$ lines~\cite{sat1,sat2,sat3}. In addition to the energy calibration, 
the time dependent instability was corrected for each SDD. The energy calibration and the 
instability correction were performed every few hours.
Among the 144 SDDs, 94 were selected for further analysis 
based on energy resolution, stability, and peak shapes.

\begin{figure}
\begin{center}
\includegraphics[width=0.6\linewidth]{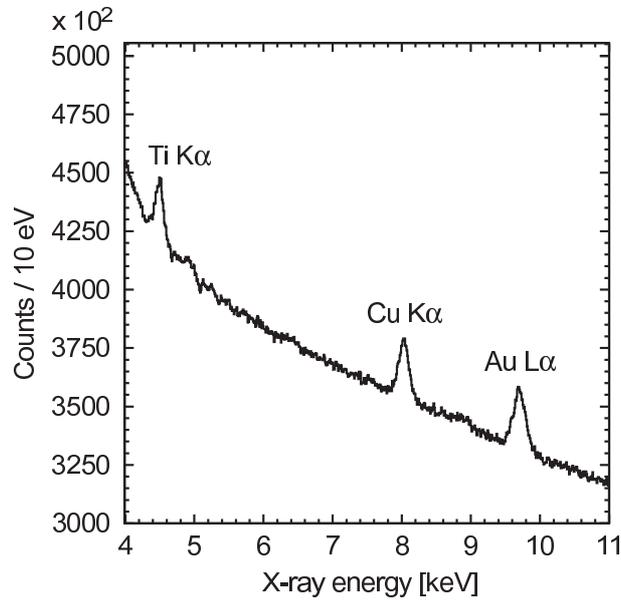}
 \caption{Energy spectrum for the selected 94 SDDs, where the timing 
uncorrelated to the $K^+K^-$ pair productions was selected. 
The accuracy of the energy scale was evaluated using the peak positions of 
the Ti K$\alpha$, Cu K$\alpha$, and Au L$\alpha$ lines.}
\label{self}
\end{center}
\end{figure}

Figure~\ref{self} shows the energy spectrum obtained with 94 SDDs with 
timing uncorrelated to the $K^+K^-$ pairs.
Together with the large beam background, there are fluorescence X-ray peaks 
at 4.5, 8.0, and 9.6 keV, which were identified as the Ti K$\alpha$, Cu K$\alpha$ and 
Au L$\alpha$ lines. The Ti and Cu lines were produced in the foils installed on
the top of the target cell, while the Au line was produced in the material of 
the printed-circuit boards of the SDDs.

The accuracy of the energy scale determined was examined using these peak positions.
The fit of the X-ray peaks showed that the gain was slightly shifted by about 6 eV.
The gain shift was expected because the hit rates of the SDDs were much higher 
in the calibration data. 
This gain shift was corrected. The accuracy of the energy scale was found to be $\pm 4$ eV, 
which was determined using the difference between the known X-ray energies and 
the peak positions after the gain correction \cite{sidd-khe3,sidd-khewidth}. 

High radiation caused by the large beam loss introduces a latch-up phenomenon in some of the SDDs.
The latch-up was automatically recovered by re-applying the voltages on the SDDs.
However, the SDDs could not measure in the periods of the latch-up.
The evaluation of X-ray event loss caused by the latch-up  (i.e. dead time correction) 
is not simple, because it depends not only on the number of latched-up SDDs, but also on the beam conditions, etc.
To dismiss the dead time correction, we selected only the periods
of data taking without any latched-up SDDs. About 20\% of the total data were rejected by this selection.

\begin{figure}
\begin{center}
\includegraphics[width=0.6\linewidth]{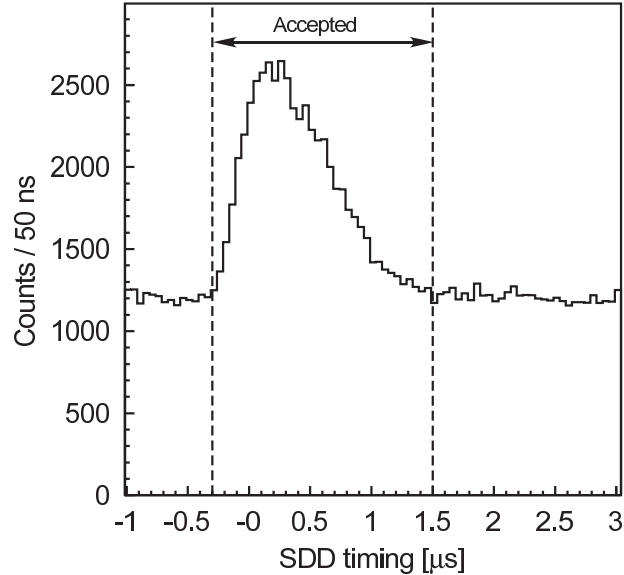}
  \caption{Timing spectrum of SDDs. 
The time difference between the $K^+K^-$ coincidence and SDD X-ray hits
is plotted. The peak region corresponds to the coincidence of the $K^+K^-$ and X-ray events.
The region indicated by arrows was selected as the good timing of the X-ray events.}
  \label{timing}
\end{center}
\end{figure}

The time differences between the coincidences in the kaon detector and the X-ray hits in the SDDs were
also measured. Figure~\ref{timing} shows the spectrum of the time difference, 
where the time walk effect was corrected using the correlation between the X-ray energy and timing. 
The origin of the time scale  is arbitrary. The continuous flat background seen 
in the figure corresponds to the beam background uncorrelated to the kaon timing. 
The events correlated to the kaon timing are seen as the peak.
The broadening of the peak corresponds to the drift time of the charges generated by the X-rays
in the SDDs. The peak area corresponds  to kaonic atom X-ray events together with 
the beam background correlated to the kaon timing (e.g., secondary particles produced by kaon decay and 
nuclear reactions). 
For the X-ray energy spectra, event selection using timing suppresses 
the background uncorrelated with the kaon timing. The timing region was selected as 
indicated by the arrows in the figure, where the X-ray event loss caused by this selection was 
determined to be less than 1\%.

\begin{figure}
\begin{center}
\includegraphics[width=0.6\linewidth]{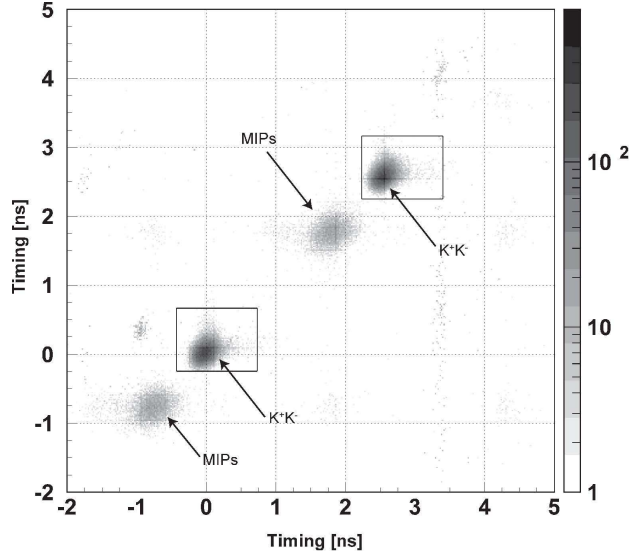}
 \caption{Timing spectrum of the kaon detector. The correlation of 
the time differences measured in the two scintillators is shown. The coincidence events 
of $K^+K^-$'s and MIPs were marked in the figure. The regions with a rectangle
were accepted as the timing windows of the $K^+K^-$ coincidence.}
 \label{kdtiming}
\end{center}
\end{figure}

The $K^+K^-$ pairs produced by $\phi$ decay were detected by the kaon detector.
The charged kaons were identified by a time-of-flight technique, using the
beam timing signals delivered by DA$\Phi$NE. 
The time differences between the beam timing signals and the coincidences in the kaon detector
were measured. Figure~\ref{kdtiming} shows a correlation of the time differences measured
in the two scintillators of the kaon detector. 
The origin of the timing is arbitrary.
The events corresponding to the $K^+K^-$ pairs were marked in the figure. 
In addition to the $K^+K^-$ pairs, fast minimum-ionizing particles (MIPs),
generated by the beam background, also passed through the two scintillators in coincidence, 
as shown in Fig.~\ref{kdtiming}. 
These $K^+K^-$ pairs and MIPs appear in two regions because the half frequency of
the beam timing signals was used. The regions marked with a rectangle were accepted 
as the timing windows for the $K^+K^-$ coincidence. The number of  $K^+K^-$ events
was determined to be
\begin{equation}
N_{trig}^{EXP} = 8.495 \cdot 10^6.
\label{exp-trig}
\end{equation}

\begin{figure}
\begin{center}
\includegraphics[width=1.0\linewidth]{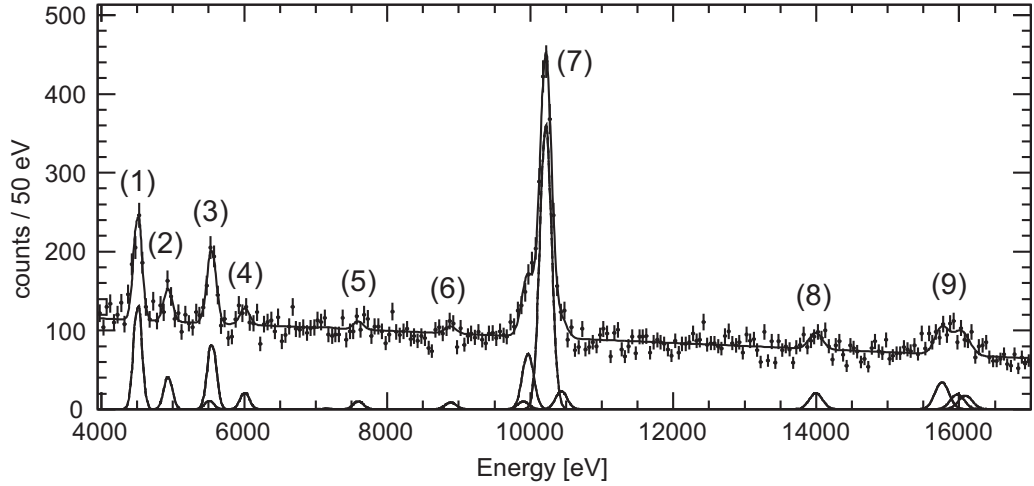}
  \caption{Energy spectrum of the kaonic atom X-rays produced in Kapton, where 
the target cell was filled with deuterium. (1) Ti K$\alpha$, (2) Ti K$\beta$, (3) 
kaonic carbon ($K^-$C) $8 \to 6$ and $6 \to 5$, (4) kaonic oxygen ($K^-$O) $7\to6$, (5)
kaonic nitrogen ($K^-$N) $6\to5$, (6)$K^-$N $7\to5$, (7)$K^-$O $8\to6$ and $6\to5$, $K^-$C $5\to4$, 
and kaonic aluminum ($K^-$Al) $8\to7$, (8)$K^-$N $5\to4$, and
(9) $K^-$C $6\to4$,  $K^-$O $7\to5$, and $K^-$Al $7\to6$ transitions.}
  \label{spect}
\end{center}
\end{figure}

With timing cuts on the SDDs and event selection on the $K^+K^-$ pairs in the kaon detector, 
a clean energy spectrum of the kaonic atom X-rays was extracted. 
Figure~\ref{spect} shows the energy spectrum from the deuterium gas target, where the number of 
SDDs used in the analysis is
\begin{equation}
N^{EXP}_{SDD} = 94.
\end{equation}

The peaks shown in the figure were attributed to X-ray transitions in the kaonic atoms formed in Kapton.
Since Kapton contains carbon, nitrogen, oxygen, and hydrogen, the X-ray transitions of 
kaonic carbon, nitrogen, oxygen were expected, whereas the kaonic hydrogen X-rays produced in Kapton 
were negligible  because of the low X-ray yields. The largest peak at around 10 keV was
identified as the kaonic carbon ($K^-$C) $5 \to 4$ transition. Other observed transitions are 
summarized in the captions of Fig. \ref{spect} and Table \ref{tab1}. 

\begin{table}
\begin{center}
\caption{List of the X-ray transitions of the kaonic atoms produced in Kapton, where
$K^-$C, $K^-$N, $K^-$O, and $K^-$Al denote kaonic carbon, kaonic nitrogen, kaonic oxygen, 
and kaonic aluminum, respectively. Because the fluorescence Ti K$\alpha$ and Ti K$\beta$ lines were
also observed, these values are also given as reference. The calculated energies of 
the transitions are shown in the second column. The number $N_X^{EXP}$ of  X-rays observed 
in a peak of the kaonic atoms is given in the last column.}
\label{tab1}
\begin{tabular}{ccc}
\hline
\hline
Transition &  Energy & Number of\\
           &  [eV]   & events ($N_X^{EXP}$) \\
\hline
$K^-$C $5\to4$   & 10216.5& $1457\pm47$\\
$K^-$C $6\to5$   &  5544.9& $ 265\pm69$ \\
\hline
$K^-$C $6\to4$   & 15759.4& $167\pm29$\\
$K^-$C $7\to5$   &  8885.8& $ 34\pm24$\\
$K^-$C $8\to6$   &  5509.6& $ 34\pm67$\\
\hline
$K^-$N $5\to4$   &13995.9 & $ 95\pm25$\\
$K^-$N $6\to5$   & 7595.4 & $ 36\pm24$\\
\hline
$K^-$O $6\to5$   &  9968.7& $280\pm51$\\
$K^-$O $7\to6$   &  6006.8& $ 69\pm24$\\
\hline
$K^-$O $7\to5$   &15973.3 & $ 92\pm38$\\
$K^-$O $8\to6$   & 9902.7 & $ 39\pm48$\\
\hline
$K^-$Al $7\to6$  &16088.3 & $ 85\pm37$\\
$K^-$Al $8\to7$  &10435.1 & $ 94\pm27$\\
\hline
Ti K$\alpha$ &  4508.9    & $392\pm31$\\
Ti K$\beta$  &  4931.8    & $124\pm25$\\
\hline
\hline
\end{tabular}
\end{center}
\end{table}

The kaonic aluminum ($K^-$Al) lines originated from kaons stopping in the solid aluminum of the setup.
Since it was very difficult to estimate the position of the sources of these X-rays, 
the kaonic aluminum yields were not evaluated. Ti fluorescence X-rays at 4.5 and 4.9 keV were 
produced in the Ti calibration foil and were correlated to the timing of the $K^+K^-$ pairs.
The contribution of kaonic deuterium X-rays produced in the gas target was negligible 
due to the low yield and expected broad line width~\cite{kd}. 

The shift and broadening caused by the strong interaction are negligible in the observed peaks.
Thus, the X-ray energies can be calculated based on the electromagnetic interaction only,
as given in Table \ref{tab1}. In the calculations, the first order of the vacuum polarization, 
the relativistic effect, the recoil correction, and the first order of the nuclear size effect are included. 
The accuracy of the calculations is estimated to be below 1 eV, which is much smaller 
than our estimated systematic error ($\pm 4$ eV). 
We can safely assume that the observed kaonic atoms 
consist of the isotopes of $^{12}$C, $^{14}$N, $^{16}$O or $^{27}$Al, because 
the natural abundances of $^{13}$C (1.1\%), $^{15}$N (0.37\%), or $^{17,18}$O (0.24\%) 
are small.

The number of X-rays observed in a peak area was obtained from the fit of the energy spectrum.
Each peak was fitted with a Gaussian function with a free intensity parameter ($I$):
\begin{equation}
G(E)=\frac{I}{\sqrt{2 \pi} \sigma}  \exp{\left(\frac{-(E-E_0)^2}{2\sigma^2} \right),}
\end{equation}
where the peak position ($E_0$) was fixed to the calculated value given in Table~\ref{tab1}.
$\sigma$ gives the Gaussian width, which represents the detector resolution.
The energy dependent function of $\sigma(E)=\sqrt{a+bE}$ was used, where 
the parameters $a$ and $b$ were determined using the Ti, Cu, and Au X-ray peaks
observed in the spectrum uncorrelated to the kaon timing. The detailed method 
is 
given in Ref.~\cite{sidd-khewidth}. 
The measured number of X-rays ($N_X^{EXP}$) was calculated as $N_X^{EXP} = I/B$, 
where $B$ is the bin width in the spectrum ($B = 50$ eV in our case).
The number of X-rays observed in each peak area is summarized in Table~\ref{tab1}. 
The peak fit functions after the background subtraction are shown in Fig.~\ref{spect}.

The X-ray detection efficiency ($\epsilon^{EXP}$) is defined as the number 
of measured X-rays ($N_X^{EXP}$) normalized by the number of $K^+K^-$ events ($N^{EXP}_{trig}$) 
recorded in the kaon detector (the kaon trigger):
\begin{equation}
\epsilon^{EXP} = \frac{N_X^{EXP}}{N^{EXP}_{trig}} 
\label{eq:exp-def}
\end{equation}
Comparing these efficiencies to those given by  Monte Carlo simulations, the absolute yields
of the kaonic atom X-rays can be determined.

\section{Monte Carlo simulations}
\label{montecarlo}
In the SIDDHARTA experiment, the positions of the kaons stopped in the setup were not measured.
Monte Carlo simulations are needed to evaluate the number of kaons stopped in the 
Kapton windows, as well as the X-ray detection efficiency of the SDDs, and the X-ray 
attenuation in Kapton and the target gas. The Monte Carlo simulations were performed
based on the GEANT4 toolkit \cite{geant4}, version 4.9.4.
All the materials and geometries used in the experiment were included in the simulations.

\begin{figure}
\begin{center}
\resizebox{1.05\columnwidth}{!}{%
 \includegraphics{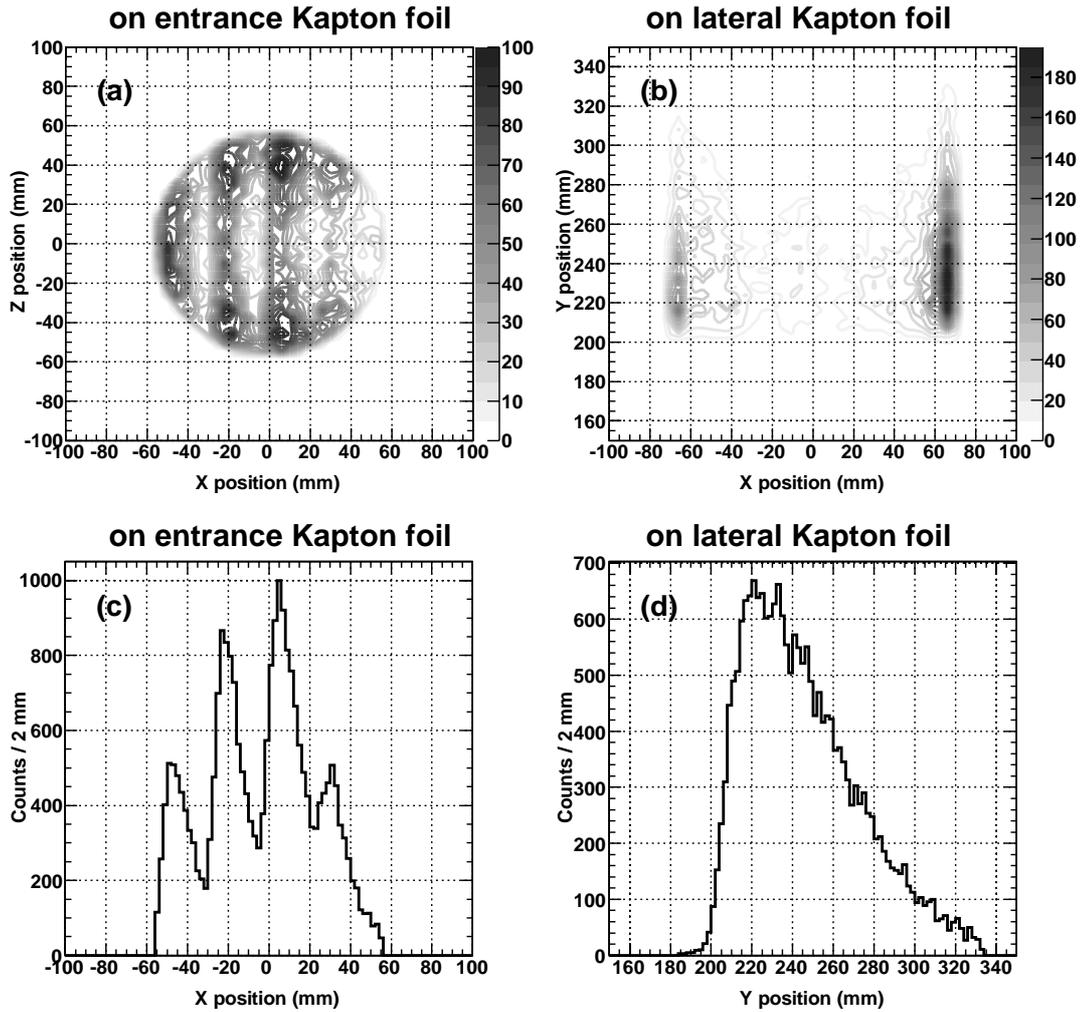} }
 \caption{The simulated distributions of the $K^-$'s stopped in 
the (a) entrance and (b) lateral windows made of Kapton. 
The $x$ axis is set to the direction of the $\phi$ boost.
The $z$ axis is parallel to the direction of the beams, and the $y$ axis gives
the vertical direction in the setup.  (c) and (d) show  the projected histograms 
onto  (a) the $x$ axis and (b) the $y$ axis, respectively.}
\label{fig:stopkpos}
\end{center}
\end{figure}

The simulations start with the production of the charged kaon pairs ($K^+K^-$) at the interaction point
(See Fig.~\ref{fig:Setup}). Although $\phi$ mesons have decay modes other than 
 $\phi \to K^+K^-$, they were not needed for the determination of the 
X-ray yields because the $K^+K^-$ pairs were efficiently detected experimentally.

The 510 MeV beams of electrons and positrons collide
with a crossing angle of 50 mrad (25 mrad per beam) 
at the interaction point. 
Thus, the $\phi$ mesons produced have a small transverse momentum of 25.48 MeV/$c$ along the horizontal plane of the colliding beams ($\phi$-boost).
The momentum distribution of the $K^+K^-$ pairs caused by the $\phi$-boost was taken into account
in the simulations. The direction of the boost, which is toward the center 
of the DA$\rm{\Phi}$NE rings, is given by the horizontal direction $x$ in our notation.

The production points of the $\phi$ mesons in DA$\Phi$NE have a finite volume,
because of the size and crossing region of the electron and positron beams.
The initial distribution of the positions of the $K^+K^-$ pairs 
was set to this volume using three-dimensional Gaussians.
The broadenings of the Gaussians
are as follows: $\sigma_x=0.26$ mm, $\sigma_y=3.2$ $\mu$m, and $\sigma_z=10.4$ mm,
where $x,y$ and $z$ give the horizontal, vertical, and beam-axis directions, respectively.
The polar angular distribution of the $K^+K^-$ pairs, which is given as
$\rm{d}\sigma/\rm{d}\cos{\theta}\sim \sin^2{\theta}$, was taken into account in the simulation. 

The interaction processes implemented for kaons were as follows: the decay, elastic scattering, 
inelastic scattering, and at-rest $K^-$ absorption. 
The photon interactions implemented were 
the photoelectric effect, Compton and Rayleigh scattering, gamma conversion, bremsstrahlung, 
and annihilation.
In addition, Auger electrons and fluorescence X-rays were produced in the de-excitations of atoms.

13 million  $K^+ K^-$ pairs were generated at the interaction point in the simulations, and
1.25 million were recorded in the kaon detector as coincidences.
The number of kaon triggers, defined as the number of $K^+K^-$ events recorded in the kaon detector, is then:
\begin{equation}
N_{trig}^{MC} = 1.250 \cdot 10^6.
\end{equation}

The 1.42 g/cm$^3$ density Kapton foils (C$_{22}$H$_{10}$N$_{2}$O$_{5}$) were used as a part of 
the cylindrical target cell.
The foils were placed at two positions: bottom (kaon entrance) and lateral. 
The number of  $K^-$'s stopped in the entrance and lateral windows was
24538 and 20858, respectively. The distributions of the $K^{-}$'s stopped 
in the Kapton windows are plotted in Fig.~\ref{fig:stopkpos}, where the origin 
of the axes was set to the position of the interaction point.

Figure~\ref{fig:stopkpos}(a) shows the distribution of the $K^-$'s stopped in the entrance window.
The $z$ axis is parallel to the direction of the beams. The direction of 
the $\phi$ boost is toward the positive sign of the $x$ axis. The gray scale is 
proportional to the number of stopped kaons, as indicated in the right side of the figure. 
The projection of the distribution onto the $x$ axis is shown in Fig.~\ref{fig:stopkpos}(c).
The peak structures seen in Fig.~\ref{fig:stopkpos}(c) are due to the use of a degrader having
a step-like shape with 2-cm intervals, which was
designed to optimize the number of $K^-$'s stopping in the gaseous target.
The disk-like distribution in Fig.~\ref{fig:stopkpos}(a) comes from the shape 
of the entrance window, where the distribution of stopped $K^-$'s is not uniform.
This non-uniformity is due to a combination of the angular dependence of 
the initial kaon momenta, of the boost effect, and of the shape of the degrader.

The distribution of the $K^-$'s stopped in the lateral window is shown in Fig.~\ref{fig:stopkpos}(b),
where the $y$ axis gives the vertical direction in the setup. The projection of 
the distribution onto the $y$ axis is shown in Fig.~\ref{fig:stopkpos}(d).
The number of  $K^-$'s stopped in the boost direction is higher, 
and the number of $K^-$'s stopped in a lower position is also higher.

\begin{table}
\begin{center}
\caption{The predicted number of X-rays detected by 94 SDDs in the simulations, where 
${ent.}$ and ${lat.}$  denote the origin of the generated X-rays:
the entrance (${ent.}$) and lateral (${lat.}$) windows.
The number of triggers is $N^{MC}_{trig} = 1.25 \cdot 10^6$.
$f_{a}$ is the atomic percentage of each element in Kapton.}
\label{table:detxray}
\begin{tabular}{crrrr}\\
\hline
\hline
Transition & Energy [keV] &  $N^{MC}_{ent.}$ & $N^{MC}_{lat.}$  & $f_{a}\ \ $\\
\hline
$K^-$C $5\to4$   & 10.2165 & 501 & 738 & 22/39\\
$K^-$C $6\to5$   &  5.5449 & 385 & 600 & 22/39\\
\hline              
$K^-$C $6\to4$   & 15.7594 & 428 & 601 & 22/39\\ 
$K^-$C $7\to5$   &  8.8858 & 499 & 764 & 22/39\\ 
$K^-$C $8\to6$   &  5.5096 & 381 & 596 & 22/39\\
\hline                                                           
$K^-$N $5\to4$   & 13.9959 &  33 &  65 & 2/39\\
$K^-$N $6\to5$   &  7.5954 &  47 &  67 & 2/39\\
\hline                       
$K^-$O $6\to5$   &  9.9687 & 122 & 167 & 5/39\\ 
$K^-$O $7\to6$   &  6.0068 &  96 & 147 & 5/39\\
\hline                                                           
$K^-$O $7\to5$   & 15.9733 &  97 & 130 & 5/39\\
$K^-$O $8\to6$   &  9.9027 & 117 & 160 & 5/39\\
\hline
\hline
\end{tabular}
\end{center}
\end{table}

\begin{table}
\begin{center}
\caption{Calculated detection efficiencies ($\epsilon^{MC}$) of 
the kaonic atom X-rays produced in Kapton. The systematic errors were 
calculated using the uncertainties
of the position of the experimental apparatus ($\pm 1$cm) and 
the thicknesses of the degrader ($\pm 50 \mu$m).
The total errors of the efficiencies, which were obtained by the combination
of the statistical and systematic errors, are given in the last column.}
\label{table:deteff}
\begin{tabular}{crcccccc}
\hline
\hline
Transition & Energy &  Efficiency        & Stat. error  & \multicolumn{2}{c}{Syst. error} & \multicolumn{2}{c}{Total error} \\
           &   [keV] &   [\%]            & [\%]         & \multicolumn{2}{c}{[\%]} & \multicolumn{2}{c}{[\%]} \\
\hline
$K^-$C $5\to4$   & 10.2165 & 0.1760 & $\pm 0.0040$ &  +0.0407 & -0.0106& +0.0409 & -0.0113 \\
$K^-$C $6\to5$   &  5.5449 & 0.1398 & $\pm 0.0036$ &  +0.0248 & -0.0126& +0.0251 & -0.0131 \\
\hline             
$K^-$C $6\to4$   & 15.7594 & 0.1462 & $\pm 0.0037$ &  +0.0135 & -0.0201& +0.0140 & -0.0205 \\
$K^-$C $7\to5$   &  8.8858 & 0.1792 & $\pm 0.0041$ &  +0.0318 & -0.0192& +0.0321 & -0.0196 \\ 
$K^-$C $8\to6$   &  5.5096 & 0.1387 & $\pm 0.0036$ &  +0.0253 & -0.0119& +0.0256 & -0.0125 \\
\hline                                                           
$K^-$N $5\to4$   & 13.9959 & 0.1558 & $\pm 0.0126$ &  +0.0316 & -0.0046& +0.0341 & -0.0134 \\
$K^-$N $6\to5$   &  7.5954 & 0.1782 & $\pm 0.0135$ &  +0.0190 & -0.0278& +0.0233 & -0.0309 \\
\hline                                                           
$K^-$O $6\to5$   &  9.9687 & 0.1812 & $\pm 0.0086$ &  +0.0326 & -0.0165& +0.0337 & -0.0186 \\ 
$K^-$O $7\to6$   &  6.0068 & 0.1523 & $\pm 0.0079$ &  +0.0207 & -0.0191& +0.0222 & -0.0207 \\
\hline                                                           
$K^-$O $7\to5$   & 15.9733 & 0.1421 & $\pm 0.0076$ &  +0.0149 & -0.0195& +0.0168 & -0.0210 \\
$K^-$O $8\to6$   &  9.9027 & 0.1735 & $\pm 0.0084$ &  +0.0448 & -0.0089& +0.0456 & -0.0123 \\
\hline
\hline
\end{tabular}
\end{center}
\end{table}

\begin{figure}
\begin{center}
\resizebox{1.0\columnwidth}{!}{%
 \includegraphics{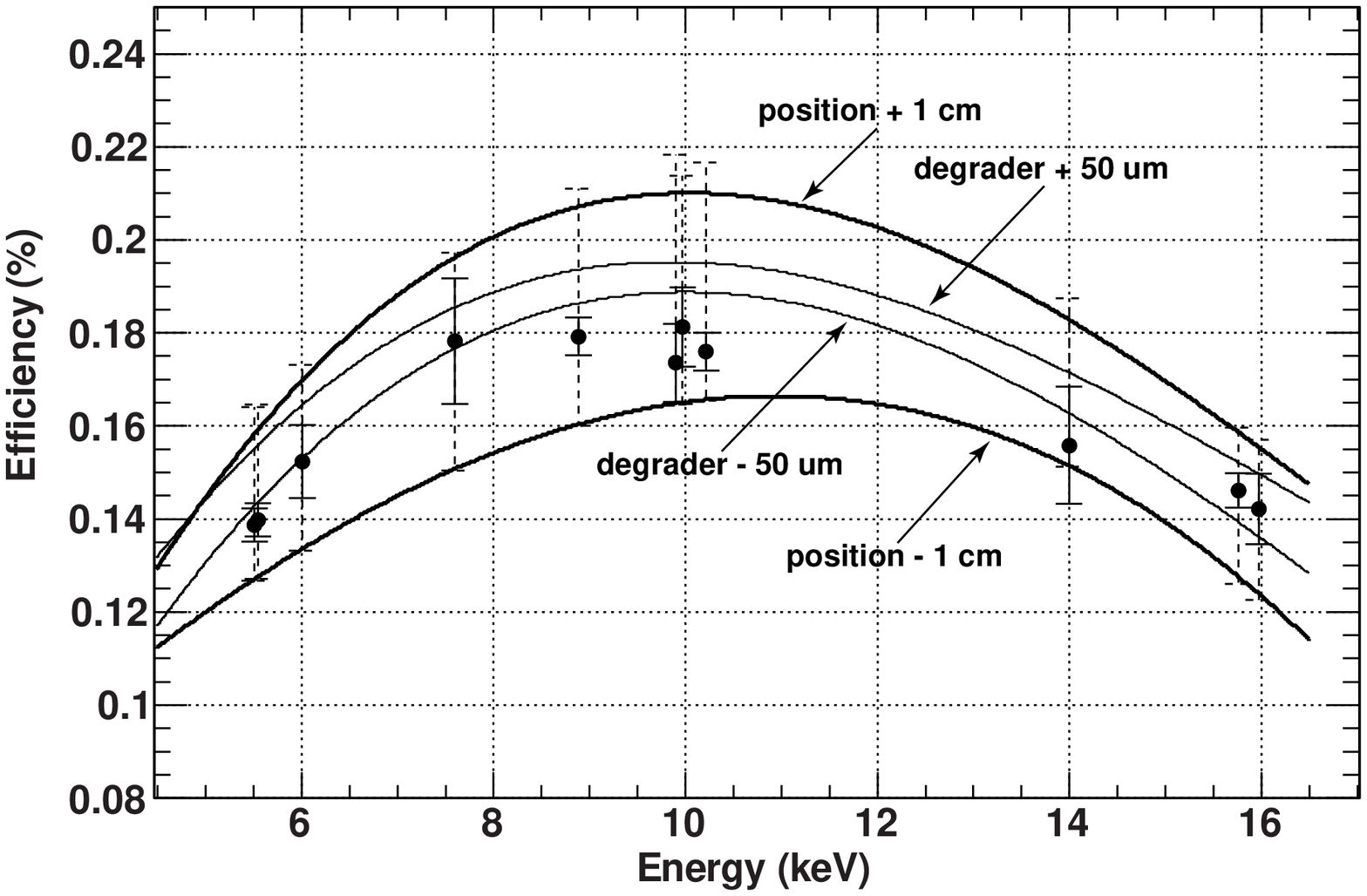} }
 \caption{Detection efficiencies ($\epsilon^{MC}$) determined by the simulations.
A black circle gives the central value of the efficiency. A solid-line error bar gives the
statistical error in the simulation.  The efficiencies, when the distance between
the interaction point of DA$\Phi$NE and the experimental apparatus were changed by $\pm 1$cm, are shown
as the curves indicated by arrows. The efficiencies caused by the uncertainty of the
degrader thicknesses ($\pm 50\mu$m) are also shown. A broken-line bar shows
the systematic error caused by the combination of the uncertainties of
the distance in the setup and degrader thicknesses.}
\label{fig:deteff}
\end{center}
\end{figure}

The process of modeling X-ray generation from kaonic atoms was done
in the at-rest process of the GEANT4 codes. The kaonic atom X-rays were generated 
at the position where the $K^{-}$ stopped.  A $K^-$ stopping in Kapton 
can form a kaonic atom with any of the following atoms: hydrogen (H), carbon (C), 
nitrogen (N), or oxygen (O). However, the kaon capture ratios of H:C:N:O 
in the kaonic atoms produced in Kapton are not well known. In the simulations, 
realistic capture ratios of kaonic atoms were not included. Instead, 
the ratios were set to be proportional to the atomic ratios in Kapton: {\it e.g.,} 
$f_{a}=22/(22+10+2+5)$ for kaonic carbon. 
The X-rays were isotropically emitted with a 100\% yield for each transition and each kaonic atom,
where no cascade calculations were involved.
The X-ray energies were fixed to  values measured experimentally.

Table~\ref{table:detxray} shows the predicted number of X-rays recorded in the 94 SDDs:
\begin{equation}
N^{MC}_{SDD} = 94,
\end{equation}
when the X-ray absorption efficiencies of the SDDs were included.
These numbers are shown separately for the two production positions in the Kapton windows: 
(${ent.}$) for the entrance window and (${lat.}$) for the lateral window.
The atomic percentages of the elements in Kapton ($f_{a}$) are also given in the table.
The values of $N^{MC}_{ent.}$ and $N^{MC}_{lat.}$  are related to $f_a$.
Because of the small $f_a$ for kaonic nitrogen, the number of kaonic nitrogen X-rays generated is small.

The X-rays produced in the entrance window and the X-rays produced in 
the lateral window cannot be distinguished experimentally.
Thus, the sum of the two values, $N^{MC}_{ent.}$ and $N^{MC}_{lat.}$,
were used for the comparison with the experimental data.
The efficiency in the simulation is defined similarly to Eq.~(\ref{eq:exp-def}):
\begin{equation}	
\epsilon^{MC} = \frac{N^{MC}_{X}}{N^{MC}_{trig}} \cdot \frac{1}{f_{a}}
= \frac{N^{MC}_{ent.} + N^{MC}_{lat.}}{N^{MC}_{trig}} \cdot \frac{1}{f_{a}},
\end{equation}
where $N^{MC}_{X}$ is the sum of $N^{MC}_{ent.}$ and $N^{MC}_{lat.}$.
The term $1/ f_{a}$ is the factor to normalize with the Kapton compound.
Table.~\ref{table:deteff} shows the efficiencies ($\epsilon^{MC}$), where the statistical errors 
are related to the number of $K^+K^-$ pairs generated in the simulation.
Figure~\ref{fig:deteff} shows the detection efficiencies ($\epsilon^{MS}$) as a function of X-ray energy.
The positions of the black circles show the central values of the efficiencies. 
The error bars shown as solid lines correspond to the sizes of the statistical errors. 
The efficiency decreases at lower energies due to higher X-ray attenuation in
the Kapton foil (and the target gas) while it decreases at higher energies
due to lower X-ray detection efficiencies of the SDDs.

The uncertainties of the parameters used in the simulation give systematic errors
in the determination of the efficiencies ($\epsilon^{MC}$). The dominant contribution was 
found to be the uncertainty in the vertical distance between the interaction point 
of DA$\Phi$NE and the experimental apparatus,
which is estimated to be about $\pm$1 cm. The contribution to the efficiencies due to this 
uncertainty is shown in Fig.~\ref{fig:deteff}, where
the curves were determined by a fit to a cubic function.
The uncertainty of the degrader thicknesses (about $\pm$50 $\mu$m in Mylar) also contributes to 
the systematic error, as shown in Fig.~\ref{fig:deteff}.
Possible other contributions (such as the uncertainties of 
the beam energy, and the region and angle of the $e^+e^-$ colliding at 
the interaction point) are relatively small (about $\pm$0.04\%), and are neglected.
The errors due to each contribution were calculated using the differences from the central values,
and thus the errors are asymmetric. The errors due to the two contributions were 
added quadratically. The combined errors are given as 
systematic errors in Table~\ref{table:deteff}. 
These systematic errors are plotted with broken line error bars in
Fig.~\ref{fig:deteff}.
The total errors consisting of the statistical and systematic errors are
shown in the last column of Table.~\ref{table:deteff}.

\section{Results and Discussion}
\label{discussion}
The X-ray yield is defined as the X-ray intensity per stopped kaon. However, 
the positions of the stopping kaons were not measured in the SIDDHARTA experiment.
Instead, the number of stopped kaons in Kapton was evaluated in the Monte Carlo simulations.
In addition, the X-ray attenuation in the setup materials and the X-ray absorption efficiencies of the SDDs were
also calculated in the simulations. 

Comparing the efficiencies determined by the measurements ($\epsilon^{EXP}$) and simulations ($\epsilon^{MC}$), 
the X-ray yields of the kaonic atoms produced in the Kapton compound can be determined.
In the comparison, the normalization was performed using 
the number of $K^+K^-$ events detected in the kaon detector ($N_{trig}$). 
The absolute X-ray yield ($Y$) per stopped $K^{-}$ was calculated as:
\begin{equation}	
Y = \frac{\epsilon^{EXP}}{\epsilon^{MC}}
= \frac{N^{EXP}_{X} / N^{EXP}_{trig}}{N^{MC}_{X} / N^{MC}_{trig}}.
\end{equation}
The correction factor, due to the missing events caused by the data analysis and the data acquisition system,
was found to be negligible.

Great care was taken in defining X-ray yields in Kapton. When a kaon stops in Kapton 
(C$_{22}$H$_{10}$N$_{2}$O$_{5}$), 
it results in X-rays from one of kaonic carbon, nitrogen, oxygen or hydrogen. 
However, there is a lack of knowledge on which atom first captures the kaon. 
To account for this, we define the X-ray yields as the number of X-rays per stopped $K^-$ in 
C$_{22}$H$_{10}$N$_{2}$O$_{5}$.

\begin{table}
\begin{center}
\caption{X-ray yields $Y$ of the kaonic atoms produced in Kapton.
The X-ray yields $Y$ were defined as the number of X-rays per 
stopped $K^-$ in C$_{22}$H$_{10}$N$_{2}$O$_{5}$. 
The last column gives the X-ray yields $Y$  normalized by the atomic percentages $f_{a}$.
}
\label{table:kaptonyield}
\begin{tabular}{crcc}
\hline
\hline
Transition & Energy &  Yield ($Y$) & $Y/f_a$ \\
           &  [keV] &   [\%]       & [\%] \\
\hline
$K^-$C $5\to4$   & 10.2165 & $9.7^{+0.7}_{-2.3}$ & $17.3^{+1.2}_{-4.1}$ \\
$K^-$C $6\to5$   &  5.5449 & $2.2^{+0.6}_{-0.7}$ & $4.0^{+1.1}_{-1.2}$ \\
\hline                                                           
$K^-$C $6\to4$   & 15.7594 & $1.3\pm0.3$         & $2.4\pm0.5$ \\
$K^-$C $7\to5$   & 8.8858  & $0.2\pm0.2$         & $0.4\pm0.3$ \\
$K^-$C $8\to6$   &  5.5096 & $0.3\pm0.6$         & $0.5\pm1.0$ \\
\hline                                                           
$K^-$N $5\to4$   & 13.9959 & $0.7\pm0.2$         & $14.0^{+3.9}_{-4.8}$ \\
$K^-$N $6\to5$   &  7.5954 & $0.2\pm0.2$         & $ 4.7^{+3.2}_{-3.1}$ \\
\hline                                                           
$K^-$O $6\to5$   &  9.9687 & $1.8^{+0.4}_{-0.5}$ & $14.2^{+3.0}_{-3.7}$ \\
$K^-$O $7\to6$   &  6.0068 & $0.5\pm0.2$         & $4.2\pm1.6$ \\
\hline                                                           
$K^-$O $7\to5$   & 15.9733 & $0.8\pm0.3$         & $6.0\pm2.6$ \\
$K^-$O $8\to6$   &  9.9027 & $0.3\pm0.3$         & $2.1^{+2.5}_{-2.6}$ \\
\hline
\hline
\end{tabular}
\end{center}
\end{table}

Using the values given in Tables~\ref{tab1},~\ref{table:deteff} and Eq. (\ref{exp-trig}), 
the absolute X-ray yields of the kaonic atoms produced in the Kapton compound were calculated.
The yield of the kaonic carbon ($K^-$C) $5\to4$ transition normalized by
C$_{22}$H$_{10}$N$_{2}$O$_{5}$ was determined to be
\begin{equation}
Y(K^-{\rm C}\mbox{ } 5\to4) = 9.7^{+0.7}_{-2.3} \%,
\end{equation}
where the errors  were calculated using the statistical errors measured in the experiment 
and the total errors in the simulations. The systematic errors related to the measurements are negligible.

Table~\ref{table:kaptonyield} shows the X-ray yields of kaonic carbon ($K^-$C), 
kaonic nitrogen ($K^-$N) and kaonic oxygen ($K^-$O) produced in Kapton.
Comparing the data of the same atoms, the yield increases as $n$ of the transition 
decreases. This tendency is related to the populations of the kaons 
during the cascades in the kaonic atoms.
The relative ratios of the successive transitions in the same atoms were determined to be
\begin{eqnarray}
 \frac{ Y(K^-{\rm C}\mbox{ } 5 \to 4)}{ Y(K^-{\rm C}\mbox{ } 6 \to 5)} &=&  4.4\pm1.2 \label{eq:kc54} \\
 \frac{ Y(K^-{\rm N}\mbox{ } 5 \to 4)}{ Y(K^-{\rm N}\mbox{ } 6 \to 5)} &=&  3.0\pm2.1 \label{eq:kn54} \\
 \frac{ Y(K^-{\rm O}\mbox{ } 6 \to 5)}{ Y(K^-{\rm O}\mbox{ } 7 \to 6)} &=&  3.4\pm1.4 \label{eq:ko65}
\end{eqnarray} 
In the calculation of the relative yields, the systematic errors in the simulations were mainly canceled
in our case. The dominant contributions to the errors were found to be the statistical errors, 
and thus the systematic errors are negligible, resulting in the errors 
given in Eqs. (\ref{eq:kc54})-(\ref{eq:ko65}) being smaller than 
the errors calculated using the values given in Table~\ref{table:kaptonyield}.
These values are larger compared to the data given in Ref.~\cite{Wiegand-Godfrey74}.
However, it is not easy to discuss a quantitative comparison, because
these ratios have a strong $Z$ dependence, as seen in Fig.~9 of ~\cite{Wiegand-Godfrey74}.

The X-ray yields $Y$ of kaonic nitrogen are smaller because of the small atomic percentage
of nitrogen in Kapton ($f_{a}=2/39$), as expected. The ratios of C:N:O in 
the X-ray yields are expected to be related to the atomic ratios of Kapton. The X-ray yields 
normalized by the atomic percentages $f_{a}$ are shown in Table~\ref{table:kaptonyield}.
The normalized yields $Y/f_a$ of kaonic carbon and kaonic nitrogen are 
almost the same within the errors both in the $5 \to 4$ and $6 \to 5$ transitions, 
while $Y/f_a$ of the kaonic oxygen transitions are higher than those
of kaonic carbon or kaonic nitrogen both in the $6 \to 5$ and $7 \to 5$ transitions.
The relative ratios of the X-ray transition yields were determined to be
\begin{eqnarray}
 \frac{ Y(K^-{\rm C}\mbox{ }5 \to 4) / f_a(\rm{C})}{ Y(K^-{\rm N}\mbox{ }5 \to 4) / f_a(\rm{N})}  
  &=&  1.23\pm0.35\\
 \frac{ Y(K^-{\rm C}\mbox{ }6 \to 5)/f_a(\rm{C})}{ Y(K^-{\rm N}\mbox{ }6 \to 5)/f_a(\rm{N})} 
  &=&  0.84\pm0.60\\
 \frac{ Y(K^-{\rm C}\mbox{ }6 \to 5)/f_a(\rm{C})}{ Y(K^-{\rm O}\mbox{ }6 \to 5)/f_a(\rm{O})} 
  &=&  0.28\pm0.09
\end{eqnarray}
The systematic errors in the simulations were found to be negligible in the calculation of the relative yields.

The yield patterns and the yield ratios can be related to the 
capture processes in the Kapton compound and the cascade processes of the kaonic atoms. 
The capture processes have been studied with several models such as the Fermi-Teller law~\cite{FT}, 
its improved models~\cite{Schneuwly,Krumshtein,Daniel,Vogel,Petruhkin} and the so-called 
meso-molecular models~\cite{Schneuwly,Ponomarev}. See also Ref.~\cite{Stanislaus}.
The hydrogen transfer process could also contribute~\cite{PLB-WG,Wiegand-Lum,Jackson}. 
Since the X-ray yields are  related to the cascade processes in addition to the capture processes, 
a detailed theoretical calculation is needed to understand the yield 
patterns determined in this experiment.

\section{Conclusions}

The kaonic atom X-rays produced in the compound of Kapton ($\rm{C}_{22}\rm{H}_{10}\rm{N}_{2}\rm{O}_{5}$) 
were measured by the SIDDHARTA collaboration. This is the first measurement 
of these X-rays using a target made of such a complex compound.
Compared to the number of X-rays generated by the Monte Carlo simulations, 
the absolute X-ray yields of the kaonic atoms produced in Kapton 
with low atomic numbers $Z$ and high $n$ transitions were determined.
We also determined the relative ratios of the successive transitions in the same atoms
and the relative ratios of C:N and C:O for the same transitions.
These results provide very important information for the theoretical studies of the capture 
and cascade processes of kaonic atoms in a compound material.

The deduced X-ray yields are also important for the measurement of kaonic deuterium X-rays
in the SIDDHARTA-2 experiment~\cite{kd}. As shown in Fig. 4 of Ref.~\cite{kd}, the kaonic atom X-rays
produced in Kapton will be present in the new experiment, although the intensities of the X-rays
will be reduced by a factor 20 resulting from the use of a higher density target.
One of the difficulties in the extraction of the shift and width of kaonic hydrogen X-rays 
was the unknown yields of the X-rays produced in Kapton. In the SIDDHARTA-2 experiment, 
the kaonic deuterium X-rays will be extracted using the X-ray yields determined in this article. 

\section*{Acknowledgments}
\label{sec:ack}
We thank C. Capoccia, B. Dulach, and D. Tagnani from LNF-INFN; and
H. Schneider, L. Stohwasser, and D. St\"{u}ckler from Stefan-Meyer-Institut,
for their fundamental contribution in designing and building the
SIDDHARTA setup. We thank as well the DA$\Phi$NE staff for the excellent working
conditions and permanent support. Part of this work was supported by
HadronPhysics I3 FP6 European Community program, Contract No. RII3-CT-2004-506078;
the European Community-Research Infrastructure Integrating Activity ``Study of Strongly Interacting Matter''
(HadronPhysics 2, Grant Agreement No. 227431), and HadronPhysics 3 (HP3), Contract No. 283286
under the Seventh Framework Programme of EU; Austrian Federal Ministry of Science
and Research BMBWK 650962/0001 VI/2/2009; Romanian National Authority for Scientific Research,
Contract No. 2-CeX 06-11-11/2006; the Grant-in-Aid for Specially Promoted Research (20002003), MEXT, Japan;
DFG Excellence Cluster Universe of the Technische Universit\"{a}t M\"{u}nchen; and 
the Austrian Science Fund (FWF): [P24756-N20].

\appendix
\section{Comparison with the experimental data}
\label{sec:comp}
From the X-ray yields produced in Kapton, it is not simple to 
extract the X-ray yields of kaonic carbon, nitrogen, and oxygen separately.
If we assume that the kaon capture ratios of C:H:N:O are simply proportional to 
the atomic percentages $f_{a}$ (although this assumption could be incorrect), 
the normalized yields $Y / f_{a}$ can be compared to the data taken with the 
targets made of single elements.

The normalized yields $Y /f_{a}$ of kaonic carbon were compared to the data of kaonic carbon 
taken with the solid target, where the data were given in Ref.\cite{Wiegand-Godfrey74}.
Table~\ref{table:comp-c} shows the comparison. Except for the $6\to4$ transition, 
the same transitions were not measured. Despite the simple assumption for 
the capture ratios, the yields of the $6\to4$ transition are consistent.

There are two nitrogen data sets with liquid\footnote{In Ref.~\cite{WG2},
there is no explicit description about the density of nitrogen. However, 
it is reasonable to assume that liquid nitrogen was used 
in the measurements.}~\cite{WG2} and gas \cite{KN-PLB} targets. 
Table~\ref{table:comp-n} shows the comparisons between our results and the 
nitrogen data. In the yield of the $5\to4$ transition, our result is 
consistent with the value given in Ref.~\cite{WG2}, while our results are much smaller than 
the values given by Ref.~\cite{KN-PLB}. Probably, density dependence of 
the X-ray yields could be also related to the yield patterns.

The oxygen data are not available. Instead, we compare with the data of H$_{2}$O, 
as shown in Table~\ref{table:comp-o}. Because reductions of the X-ray yields  
in the hydride materials were reported \cite{PLB-WG,Wiegand-Lum}, 
these values should be used for reference only.

\begin{table}
\begin{center}
\caption{The X-ray yields $Y$ of kaonic carbon produced in Kapton normalized by the atomic percentage $f_{a}$.
The X-ray yields shown in the third column were taken from Ref.~\cite{Wiegand-Godfrey74}.}
\label{table:comp-c}
\begin{tabular}{cccc}
\hline
\hline
Transition & $Y / f_{a}$ & Yield \\
           &    [\%]   & [\%] \cite{Wiegand-Godfrey74}\\
\hline
$K^-$C $8\to6$   & $0.5\pm1.0$ && \\

$K^-$C $6\to5$   & $4.0^{+1.1}_{-1.2}$ &&\\
$K^-$C $7\to5$   & $0.4\pm0.3$ &&\\

$K^-$C $5\to4$   & $17.3^{+1.2}_{-4.1}$ &&\\
$K^-$C $6\to4$   & $2.4\pm0.5$ &$2.9\pm1.1$&\\

$K^-$C $4\to3$   &                        &$36\pm6$&\\
$K^-$C $5\to3$   &                        &$7.0\pm1.4$&\\
$K^-$C $6\to3$   &                        &$2.8\pm0.8$&\\
$K^-$C $3\to2$   &                        &$2.8\pm0.8$&\\

\hline
\hline
\end{tabular}
\end{center}
\end{table}

\begin{table}
\begin{center}
\caption{The X-ray yields $Y$ of kaonic nitrogen produced in Kapton normalized by $f_{a}$.
The X-ray yields shown in the third column were taken from Ref.~\cite{WG2}.
In the 4th column, the data in the gas target \cite{KN-PLB} are shown.
}
\label{table:comp-n}
\begin{tabular}{cccc}
\hline
\hline
Transition & $Y / f_{a}$ & Yield & Yield \\
           &    [\%]   & [\%] \cite{WG2} & [\%] \cite{KN-PLB}\\
\hline
$K^-$N $7\to6$   &                        && $41.5\pm8.7{\rm(stat.)}\pm4.1{\rm(sys.)}$ \\
$K^-$N $6\to5$   & $ 4.7^{+3.2}_{-3.1}$ && $55.0\pm3.9{\rm(stat.)}\pm5.5{\rm(sys.)}$ \\
$K^-$N $5\to4$   & $14.0^{+3.9}_{-4.8}$ & $13\pm4$ & $57.4\pm 15.2{\rm(stat.)}\pm5.7{\rm(sys.)}$\\
$K^-$N $6\to4$   & & $4.0\pm1.7$ &  \\
$K^-$N $4\to3$   & & $38\pm6$ &  \\
$K^-$N $5\to3$   & & $3.4\pm1.0$ &  \\
$K^-$N $6\to3$   & & $2.1\pm0.7$ &  \\
$K^-$N $3\to2$   & & $2.5\pm0.9$ &  \\

\hline
\hline
\end{tabular}
\end{center}
\end{table}

\begin{table}
\begin{center}
\caption{The X-ray yields $Y$ of kaonic oxygen produced in Kapton normalized by $f_{a}$.
The X-ray yields shown in the third column were taken from Ref.~\cite{Wiegand-Godfrey74}, 
where the target was H$_{2}$O. The yield of the $6\to5$ transition in Ref.~\cite{Wiegand-Godfrey74} includes
the yield of the $8\to6$ transition.}
\label{table:comp-o}
\begin{tabular}{cccc}
\hline
\hline
Transition & $Y / f_{a}$ & Yield \\
           &    [\%]   & [\%] \cite{Wiegand-Godfrey74}\\
\hline
$K^-$O $7\to6$   & $4.2\pm1.6$ &\\
$K^-$O $8\to6$   & $2.1^{+2.5}_{-2.6}$ & (see caption) \\
$K^-$O $6\to5$   & $14.2^{+3.0}_{-3.7}$ & $27\pm10$\\
$K^-$O $7\to5$   & $6.0\pm2.6$ & $2.0\pm1.1$\\
$K^-$O $5\to4$   & & $13\pm2$ \\
$K^-$O $6\to4$   & & $3.1\pm0.7$ \\
$K^-$O $7\to4$   & & $1.4\pm0.5$ \\
$K^-$O $4\to3$   & & $15\pm2$ \\
$K^-$O $5\to3$   & & $3.4\pm0.6$ \\
$K^-$O $6\to3$   & & $2.2\pm0.4$ \\
$K^-$O $7\to3$   & & $1.4\pm0.3$ \\
$K^-$O $8\to3$   & & $0.6\pm0.2$ \\

\hline
\hline
\end{tabular}
\end{center}
\end{table}

\end{document}